# REPORTS

# Windows Through the Dusty Disks Surrounding the Youngest Low-Mass Protostellar Objects

J. Cernicharo,[1]* A. Noriega-Crespo,[2] D. Cesarsky,[3] B. Lefloch,[1,4] E. González-Alfonso,[1] F. Najarro,[1] E. Dartois,[5] S. Cabrit[6]

The formation and evolution of young low-mass stars are characterized by important processes of mass loss and accretion occurring in the innermost regions of their placentary circumstellar disks. Because of the large obscuration of these disks at optical and infrared wavelengths in the early protostellar stages (class 0 sources), they were previously detected only at radio wavelengths using interferometric techniques. We have detected with the Infrared Space Observatory the mid-infrared (mid-IR) emission associated with the class 0 protostar VLA1 in the HH1-HH2 region located in the Orion nebula. The emission arises in three wavelength windows (at 5.3, 6.6, and 7.5 micrometers) where the absorption due to ices and silicates has a local minimum that exposes the central part of the young protostellar system to mid-IR investigations. The mid-IR emission arises from a central source with a diameter of 4 astronomical units at an averaged temperature of ~700 K, deeply embedded in a dense region with a visual extinction of 80 to 100 magnitudes.

Our lack of knowledge of star formation processes led to an empirical classification of the evolutionary phases of low-mass protostars into four classes: 0, I, II, and III. These describe the amount of material available for accretion versus the mass of the central object, providing the evolutionary status of the system (*1–3*). Class 0 objects are the youngest protostars; they are surrounded by large and dusty envelopes that feed the central objects and their protoplanetary disks. These sources undergo violent ejection of matter related to accretion processes. The shockwaves created when the protostellar ejecta collides with the surrounding gas produce the Herbig-Haro (HH) jets observed at optical wavelengths. These jets seem to drive the bipolar molecular outflows (*4–6*) detected around protostars and represent a second mass loss–driven phenomenon taking place during the earliest evolutionary stages of the star formation process (7). Class 0 sources are characterized by a spectral energy distribution (SED) peaking in the far-IR, similar to a black body at 15 to 30 K, and by the presence of a highly collimated molecular and/or ionic outflow (1–3). Because of the heavy obscuration from the large amounts of circumstellar material, they have been detected and studied only at far-IR and millimeter wavelengths ($\lambda > 25$ µm). Some class 0 sources also have a compact radio continuum source detected at centimeter wavelengths. All these observational properties indirectly indicate a young stellar object. Intermediate classes (I and II), which have less mass in the placentary envelope and show a decrease of material ejection during accretion (8) relative to class 0 objects, have been studied at mid- and near-IR wavelengths. Class I sources are still embedded in copious amounts of material. Their SED rises in the mid-IR range and decreases at longer wavelengths. The ejection of matter in the surrounding gas is less violent than in class 0. Class II corresponds to T-Tauri stars; their radiation peaks in the near-IR and decreases in the mid-IR. Their SED reveals a small contribution originating from the disk in which they were formed. Class III objects are also pre–main sequence stars but have hardly any surrounding circumstellar material and show little or no IR excess.

[1]Consejo Superior de Investigaciones Científicas, Instituto de Estructura de la Materia, Departamento Física Molecular, Serrano 121, 28006 Madrid, Spain. [2]Space Infrared Telescope Facility (SIRTF) Science Center, California Institute of Technology, Pasadena, CA 91125, USA. [3]Institut d'Astrophysique Spatiale, Bât. 121, Université de Paris XI, 94500 Orsay Cedex, France. [4]Observatoire de Grenoble, Domaine Universitaire de Grenoble, 414 rue de la Piscine, 38406 St. Martin d'Hères, France. [5]Institute de Radioastronomie Millimétrique, Domaine Universitaire de Grenoble, 300 rue de la Piscine, 38406 St. Martin d'Hères, France. [6]Département d'études de la Matière en Infrarouge et Millimètrique, UMR 8540 du CNRS, Observatoire de Paris, 61 Av. de l'Observatoire, F-75014 Paris, France.

*To whom correspondence should be addressed. E-mail: cerni@astro.iem.csic.es

Models (9, 10) predict that all the processes involved in the formation of the central source and its disk, as well as in the ejection of matter in the surrounding medium, are circumscribed to a small region of a few astronomical units. However, in these early stages (class 0) the nascent low-mass protostars are so deeply embedded in their placentary cloud that they cannot be detected at optical or at near- to mid-IR wavelengths. We show that despite the enormous absorption associated with the class 0 objects, there are some mid-IR windows that permit direct viewing of the central source.

We have carried out continuum observations of the HH1-HH2 system in the mid-IR with the IR camera ISOCAM (11) onboard the Infrared Space Observatory [ISO (12)], and at millimeter wavelengths using the 19-channel bolometer array installed at the 30-m telescope of the Institute de Radioastronomie Millimétrique (IRAM). The HH1-HH2 system lies in the L1641 molecular cloud in Orion, at a distance of 460 pc (13). It exhibits a pronounced protostellar jet (14, 15) and harbors several protostellar sources at various stages of evolution. This jet has a molecular outflow counterpart (6) whose powering source was identified as the embedded protostar VLA1 (16), a possible class 0 source (17, 18). Another source, VLA2, also very young and only a few arc seconds from VLA1, drives an additional bipolar outflow (6) nearly perpendicular to that emerging from VLA1.

To study the continuum radiation, we took two maps in the circular variable filter [CVF (12)] spectral mode of ISOCAM providing at each sky pixel a full spectrum from 5 to 16.5 µm, with spectral resolution of ~40. The first map, with a pixel field of view (PFOV) of 6 arc sec, was centered on the HH2 object. In parallel with this map, we took an image in the LW2 band ($\lambda \approx 7$ µm, $\Delta\lambda \approx 4$ µm). The second CVF map, with a PFOV of 3 arc sec, was shifted to include the so-called Cohen-Schwartz (CS) star, a strong IR emitter that we used to derive the astrometry of all maps. We estimate the position of the CS star to be accurate in our maps to within 1 to 2 arc sec.

The observed composite spectrum of VLA1 and VLA2 (henceforth VLA1+VLA2) in the 5- to 17-µm range (Fig. 1) is characteristic of deeply embedded sources. The emission integrated over the 12-$\mu$m IRAS (Infrared Astronomical Satellite) band is below the detection limit of this instrument. The ISO spectrum shows saturated absorption bands of silicates, $H_2O$, $CH_3OH$, and probably $CO_2$ at 15.2 $\mu$m. The continuum radiation emitted in hot regions of the source is reabsorbed in the coldest part of the cloud by the ice mantles of dust grains. These ices are formed through the condensation of volatile species, such as of $H_2O$, $CO_2$, $CH_4$, and $CH_3OH$, onto the surface of refractory grains composed of silicates (seen at 9.7 $\mu$m in absorption) and other materials. The signal-to-noise ratio of the data is 5 for the weak continuum emission and $\sim$10 in the emission windows at 5.3, 6.6, and 7.5 $\mu$m (Fig. 1). The shape of the ISO spectrum around 16 $\mu$m could be affected by instrumental problems (transient memory effects). The emission windows in the mid-IR are found only toward the class 0 source VLA1+VLA2 and toward the class I source VLA4 (see below). Around them there is not such an emission. Moreover, the spectra do not exhibit the diffuse unidentified IR bands that have been suggested to be caused by polycyclic aromatic hydrocarbons (PAHs). No PAH emission is found in a region of 30" $\times$ 30" around VLA1.

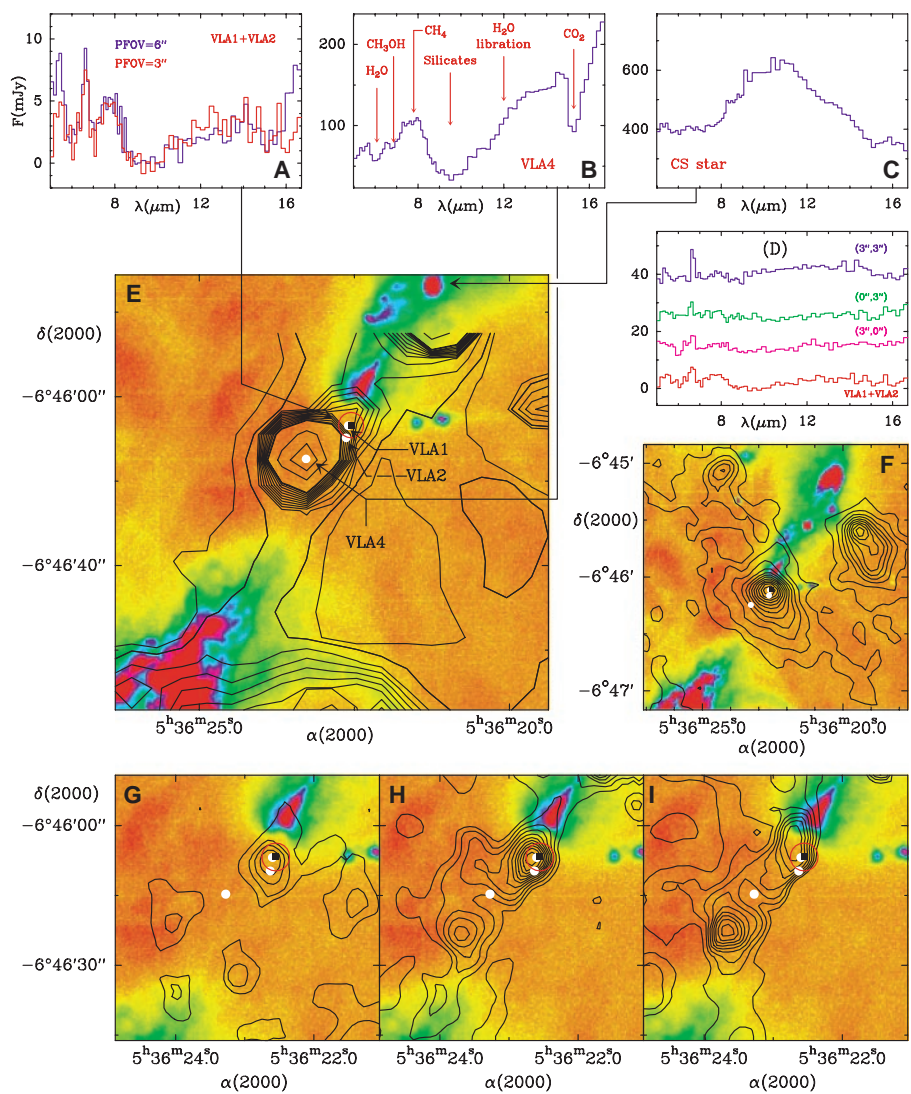

**Fig. 1.** (**A** to **C**) Observed IR emission toward VLA1+VLA2 (A), VLA4 (B), and the CS star (C). The spectra for VLA4 and the CS star correspond to those obtained with a POFV of 3 arc sec, whereas for VLA1+VLA2 we show the spectra observed in both CVFs. Note the emission through the three IR windows. (**D**) The observed IR emission around VLA1+VLA2 with a PFOV of 3 arc sec. (**E**) Optical (17) image (SII 6717/31 Å, in arbitrary units) of the central HH1-HH2 region (colors) with a superposition (contours) of the IR emission detected with the LW2 filter of ISOCAM. The positions of VLA1, VLA2, and VLA4 are indicated. The position toward which we have discovered the three IR windows is indicated by a solid square and coincides, within the astrometric errors, with that of VLA1+VLA2; the red circle around the square indicates the PFOV of the IR observations (6 arc sec). (**F**) Optical image (same as before) with the contours of the continuum emission at 1.3 mm (1300 µm as observed with the 30-m IRAM radiotelescope). The contours are between 0.04 and 0.4 Jy per beam and show the presence of a core centered on VLA1+VLA2 surrounded by an extended envelope indicating the class 0 nature of the powering engine of the HH1-HH2 system (see text). (**G** to **I**) Optical image (as above) with the corresponding IR emission in the 5.3- and 6.6-µm windows minus the adjacent continuum (G), the total continuum emission between 5 and 7 µm (H), and the total continuum emission between 13 and 14 µm (I). In these three panels, the emission of the source VLA4 has been removed by fitting the ISOCAM point spread function (PSF) to the CVF spectra.

The absorption properties of the ices in the VLA1+VLA2 spectrum could be derived from the class I protostar VLA4, a nearby strong source that was also detected in the K band (19) and at radio wavelengths (20). VLA4's CVF spectrum (Fig. 1B) is characteristic of an embedded source and is similar, within an intensity scale factor of $\sim$100, to the spectra observed toward deeply embedded objects in massive star-forming regions, in particular to the ISO spectra of NGC 7009S (21) and of NGC 7538 IRS9 (22, 23). The VLA4 ISO spectrum differs from that of VLA1+VLA2 in that the ice absorption bands are not saturated, allowing us to determine the absolute column densities ($N$) of the ice mantles' constituents. Using integrated absorption cross sections measured in the laboratory (24–26), we obtained $N(CO_2)$ = 1.9 ($\pm$0.3) $\times$ $10^{18}$ cm$^{-2}$, $N(H_2O)$ = 4.8 ($\pm$0.7) $\times$ $10^{18}$ cm$^{-2}$, and—assuming the so-called 6.85-$\mu$m band is due to methanol ice—$N(CH_3OH)$ = 4.8 ($\pm$0.5) $\times$ $10^{18}$ cm$^{-2}$ for VLA4. From these values we constructed an empirical absorption profile in which the opacity at different wavelengths scales with visual extinction $A_V$. The model also includes the absorption by silicates and by the water libration band. The general dependency of dust absorption with wavelength is assumed to follow a $\lambda^{-1}$ law. The resulting synthetic spectrum of VLA4 (Fig. 2) indicates that the embedded source should have a temperature of $\sim$500 K, a diameter of 2.5 $\times$ $10^{13}$ cm (1.7 AU), and a minimum visual absorption of 40 magnitudes. For VLA1+VLA2, the best fit with a constant temperature component corresponds to a heavily extincted ($A_V$ = 80) source with a diameter of 4.1 AU (6.2 $\times$ $10^{13}$ cm) and a temperature of $\sim$700 K (see Fig. 2). A better fit could be obtained with a larger visual absorption and two black bodies at temperatures of $\sim$400 and $\sim$1000 K, respectively. Although a temperature gradient could be expected between the central object and the external parts of its surrounding disk, the lack of spectral data below 5 $\mu$m prevents us from reaching conclusions about the presence of the hotter component (1000 K). The obtained temperatures and sizes are representative of averaged properties of the central warm region of the low-mass protostar. The luminosity of the central object of VLA1+VLA2 is 40 $L_\odot$ (where $L_\odot$ is the luminosity of the sun), in good agreement with previous observational determinations at longer wavelengths (17, 18), which give 50 $L_\odot$.

Although at mid-IR wavelengths VLA4 is much brighter than VLA1+VLA2, at millimeter wavelengths the dust emission is dominated by the latter. A map at 1250 $\mu$m shows a bright condensation, 0.036 pc in diameter, centered on VLA1+VLA2 (Fig. 1F). There is no evidence for strong emission associated with VLA4, which lies at the border of the dust envelope of VLA1+VLA2. Also, although VLA1+VLA2 is clearly associated with a bipolar molecular outflow (6), there is no evidence of outflowing gas around VLA4 (6). These differences suggest that VLA4 is more evolved than VLA1+VLA2 and is probably in the class I stage. Assuming dust parameters typical of class 0 sources (spectral index of 1.5, dust temperature of 30 K, and absorption coefficient at 250 $\mu$m of 0.1 g cm$^{-2}$), we derive for the dust envelope of VLA1+VLA2 a mass of 1.5 $M_\odot$ (where $M_\odot$ is the mass of the sun) and a hydrogen column density of 1.4 $\times$ $10^{23}$ cm$^{-2}$, corresponding to $A_V$ = 78, in good agreement with the visual absorption estimated above.

In the spectrum of VLA1+VLA2 there are three strong peaks centered at 5.3, 6.6, and 7.5 $\mu$m above the continuum. Although the continuum-subtracted emission (Fig. 1G) within these IR windows arises from an unresolved source that peaks at the position of VLA1+VLA2, the total continuum emission in the windows is clearly resolved (Fig. 1H). This is also seen in the total continuum emission between 13 and 14 $\mu$m (Fig. 1I). The emission in both "filters" has similar spatial behavior and traces some extended warm dust emission along the jet of HH1-HH2. Further, this continuum emission is also detected in the broad ISOCAM LW2 filter ($\lambda \approx$ 7 $\mu$m, $\Delta\lambda \approx$ 4 $\mu$m) as a narrow and collimated jet between VLA1+VLA2 and the HH2 object (Fig. 1E).

VLA1 is not a unique source. We found many other class 0 candidates that exhibit similar IR excesses in the ISO database. The spectra of four well-known class 0 sources (2, 9, 27) with sufficiently accurate astrometry are shown in Fig. 2. They present similar features to VLA1+VLA2, including strong absorption by ices and emission in the IR windows at 5.3, 6.6, and 7.5 $\mu$m. Although the signal-to-noise ratio of the ISO spectra in VLA4 and the other class 0 sources shown in Fig. 2 is high ($\sim$50 to 100), the last part of the spectrum around 16 $\mu$m could be affected by transient memory effects. In all cases the visual absorption must be $\sim$80 magnitudes. Most of the sources exhibit a steep increase of the IR flux at larger wavelengths (e.g., Cep

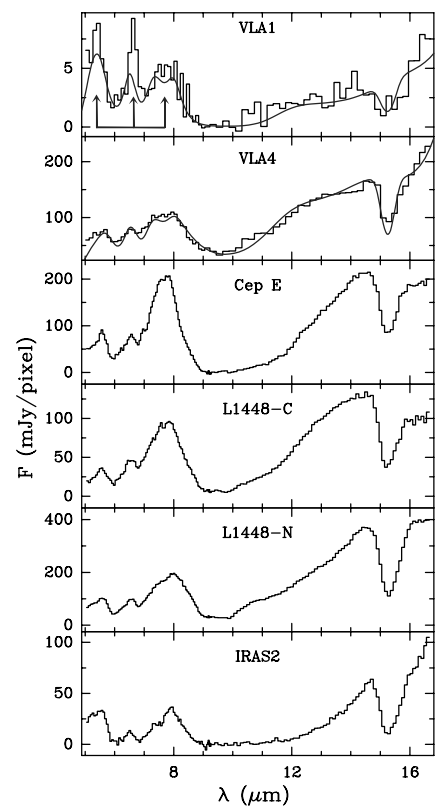

**Fig. 2.** Mid-IR spectra between 5 and 17 $\mu$m toward several class 0 sources as observed with ISOCAM. All of them show strong absorption by ices and silicates. The three spectral windows are indicated by arrows in the VLA1 panel. The data for L1448-C, L1448-N, Cep E, and IRAS2 have not been analyzed as carefully as those for VLA1+VLA2 and VLA4 in HH1-HH2, and they represent a first look at the IR emission detected by ISO. These spectra need additional refinements such as a fit of the instrumental point source function to the data. The astrometry for all the sources is better than 2 to 3 arc sec and has been derived from other bright objects in the observed fields (except for L1448-C, which is at the edge of the field of view of ISOCAM and could have a larger position uncertainty). The VLA1 fit (continuous line) is obtained from the empirical absorption derived for VLA4 (continuous line) by scaling the total visual absorption by a factor of 2 (see text).

E or IRAS2), more pronounced than toward VLA1+VLA2, which can be reproduced only by assuming a colder, optically thick emission for the central source. Hence, the observed behavior rules out the possibility of a dust jet origin of moderate column density (optically thin) for the emission in the innermost region. Nevertheless, the dusty jet is detected as an extended component in our maps (Fig. 1) and makes a contribution to the continuum emission in the 5- to 17-µm spectra observed with ISOCAM.

The large amount of ices found around these sources shows that molecular depletion onto dust grains is a major process in the cold dense regions where low-mass stars form. The profile of the ices' absorption varies from source to source (see Fig. 2), but in all of them the three mid-IR windows are present. In addition to the ices of $H_2O$, $CH_3OH$, and $CO_2$, the observed spectra toward the low-mass protostars shown in Fig. 2 indicate the presence of $CH_4$ and of another ice absorber at 7.35 µm. The amount of each component depends on the chemical evolution of the protostellar condensation. In addition, the dust in the protoplanetary disk could be heated by the radiation from the central object or from the shocks associated to the accretion processes, leading to evaporation of volatile species. The dust grains, which initially contain amorphous silicates, could be submitted to a transition phase in the evolution of the protostar region toward a stellar object and its protoplanetary disk. The temperatures we derive for the central class 0 objects are high enough to make such phase transitions possible in the innermost regions of the protoplanetary disks. In this context, we note that the spectrum toward the CS star (Fig. 1C) exhibits a broad band in emission that suggests the presence of crystalline silicates (*28*).

The good agreement between the observed luminosity at millimeter wavelengths and the predicted value for the emitting source in the few central astronomical units of VLA1+VLA2 closely relates the origin of the emission in the mid-IR and that observed at longer wavelengths from the extended cold dust. We therefore suggest that the emission detected through the three mid-IR windows traces the few inner astronomical units of the warm protoplanetary envelope (disk) around the accreting protostar, which is deeply embedded in a much larger cold core. We cannot exclude the possibility that the IR emission observed toward the sources comes from a more evolved companion; however, among the sources shown here, signs of multiple components in the protostellar core have been reported only toward L1448-C.

The present classification of protostars does not consider the range of concentrations of ices and silicates in the protostellar envelope, nor does it consider the nature of the hot central object evident in the ISO data. This classification scheme is more representative of the properties of the outer cold dust envelope than of the physical properties and processes in the innermost regions, where the star and its proto–planetary system are being formed.

# Control of Energy Transfer in Oriented Conjugated Polymer–Mesoporous Silica Composites

Thuc-Quyen Nguyen, Junjun Wu, Vinh Doan, Benjamin J. Schwartz,* Sarah H. Tolbert*

Nanoscale architecture was used to control energy transfer in semiconducting polymers embedded in the channels of oriented, hexagonal nanoporous silica. Polarized femtosecond spectroscopies show that excitations migrate unidirectionally from aggregated, randomly oriented polymer segments outside the pores to isolated, aligned polymer chains within the pores. Energy migration along the conjugated polymer backbone occurred more slowly than Förster energy transfer between polymer chains. The different intrachain and interchain energy transfer time scales explain the behavior of conjugated polymers in a range of solution environments. The results provide insights for optimizing nanostructured materials for use in optoelectronic devices.

One of the outstanding challenges in the design of nanostructured materials is the fabrication of systems that allow the flow of energy to be controlled and directed to regions useful for a desired purpose. In the photosynthetic reaction center, for example, light energy harvested at multiple sites spontaneously flows to the appropriate location in the complex so that electron transfer can take place (*1*). In attempts to mimic this type of behavior, scientists have linked chromophores with continually decreasing band gaps along polymer chains (*2*) or layered chromophores in thin film heterostructures (*3, 4*). Although good directional energy transfer can be achieved in these systems, energy is lost because large differences in the emission energy of each successive chromophore are needed.

Department of Chemistry and Biochemistry, University of California, Los Angeles, Los Angeles, CA 90095–1569, USA.

*To whom correspondence should be addressed. E-mail: tolbert@chem.ucla.edu and schwartz@chem.ucla.edu